\documentclass[12pt,preprint]{aastex}
\usepackage{natbib}
\usepackage{graphicx}
\shorttitle{A strong falsification of the universal RAR}
\begin{document}
\title{A strong falsification of the universal radial acceleration relation in galaxies}
\author{Man Ho Chan}
\affil{Department of Science and Environmental Studies, The Education University of Hong Kong, Hong Kong, China}
\email{chanmh@eduhk.hk}

\begin{abstract}
In the past few decades, many studies revealed that there exist some apparent universal relations which can describe the dynamical properties in galaxies. In particular, the radial acceleration relation (RAR) is one of the most popular relations discovered recently which can be regarded as a universal law to connect the dynamical radial acceleration with the baryonic acceleration in galaxies. This has revealed an unexpected close connection between dark matter and baryonic matter in galaxies. In this article, by following the recent robust Galactic rotation curve analyses, we derive the Galactic RAR (GRAR) and show for the first time that it deviates from the alleged universal RAR at more than $5\sigma$. This provides a strong evidence to falsify the universal nature of RAR in galaxies claimed in past studies. 
\end{abstract}

\section{Introduction}
It has been argued that there exist some apparent universal relations which can connect the dynamical properties between dark matter and baryonic matter in galaxies. For example, the Tully-Fisher relation \citep{Tully} and Faber-Jackson relation \citep{Faber} show that the luminosity of galaxies is somewhat correlated with the stellar velocities in galaxies. As the luminosity and stellar velocities are determined by baryonic content and dark matter content respectively, these relations have revealed some potential correlations between dark matter and baryonic matter, which is not expected based on the standard Lambda Cold Dark Matter model. Therefore, investigating in these universal relations is utterly important to understand the potential interplay between dark matter and baryons.

Recently, there is one intriguing relation discovered called the radial acceleration relation (RAR) \citep{McGaugh,Lelli,Mistele}, which connects the dynamical radial acceleration with baryonic acceleration in galaxies. By analyzing the data of 153 rotating galaxies in the SPARC sample, the dynamical radial acceleration $a_{\rm dyn}$ and baryonic acceleration $a_{\rm bar}$ follow a universal analytic relation with small scatters \citep{McGaugh}:
\begin{equation}
a_{\rm dyn}=\frac{a_{\rm dyn}}{1-\exp(-\sqrt{a_{\rm dyn}/a_0})},
\end{equation}
where $a_0=1.20 \pm 0.02$ (random) $\pm 0.24$ (systematic) $\times 10^{-10}$ m/s$^2$. This RAR is tantamount to a natural law for rotating galaxies as claimed in \citet{McGaugh}. Furthermore, a more recent study combining kinematic and lensing data shows that the applicability of RAR can even extend to different types of galaxies, including early- and late-type galaxies, over a large dynamic range \citep{Mistele}, which suggests the existence of a consistent and universal relation manifested in all galaxies. 

Apart from the unexpected close relation between dark matter and baryonic matter, this relation also suggests the existence of a universal acceleration scale $a_0$, which is consistent with the theory of Modified Newtonian Dynamics (MOND) \citep{Li}. Also, the universal analytic relation in Eq.~(1) is consistent with one of the interpolating functions suggested in MOND \citep{Dutton}. Therefore, the RAR becomes one important indicator to challenge the dark matter model \citep{Li2}. Nevertheless, it should be noted that some recent simulations using the Lambda Cold Dark Matter model can still reproduce the RAR \citep{Stone,Paranjape}.

On the other hand, there are some recent studies challenging the claimed universality of RAR. For example, the RAR shown in galaxies is not consistent with the data of galaxy clusters \citep{Chan,Chan2}. Also, whether the RAR can be applied in elliptical galaxies is controversial \citep{Chae,Chan3,Dabringhausen}. Generally speaking, verifying or falsifying the RAR requires very high quality measurements of rotation velocity in galaxies. Too large uncertainty in velocity measurements would make the judgment of the existence of a universal RAR very difficult \citep{Desmond}. Fortunately, some recent studies using the data of Gaia have produced accurate rotation curves in our Galaxy. The percentage errors in rotation velocities can be smaller than 5\% \citep{Eilers,Ou,Jiao}. These data are very useful in examining our understanding about dynamics in galaxies, especially in the low acceleration regime \citep{Chan4}. In this article, by using two recent robust analyses of Galactic rotation curve (GRC), we derive the Galactic RAR (GRAR) and show that it deviates significantly from the universal RAR derived from the SPARC sample. This significantly challenges the universal nature of RAR in galaxies claimed in past studies.

\section{Modeling the radial acceleration}
Recently, some studies have analyzed the new GRC data obtained by the Gaia Collaboration \citep{Eilers,Ou,Jiao}. The Gaia DR 3 has given much better improved parallaxes and proper motions measurement than that in previous measurements \citep{Jiao}. Generally speaking, the GRCs obtained by different studies are consistent with each other. In the followings, we mainly follow two most recent robust analyses of the GRC data \citep{Ou,Jiao} and obtain the dynamical radial acceleration of our Galaxy.   

The dynamical radial acceleration is given by
\begin{equation}
a_{\rm dyn}=\frac{V_c^2}{R},
\end{equation}
where $V_c$ is the stellar rotation velocity at the distance $R$ from the Galactic Center. The values of $V_c$ at different $R$ can be found in \citet{Ou,Jiao}, in which the data in \citet{Ou} cover a larger range of $R=6-27$ kpc. The dynamical radial acceleration can be directly calculated from these data.

For the baryonic radial acceleration, we need to adopt a baryonic model to calculate the value. We follow the benchmark baryonic model outlined in \citet{Misiriotis,deSalas}, in which the model parameters are determined by direct baryonic observations \citep{Misiriotis}. This model has been adopted by many recent studies \citep{Ou,Jiao}. The baryonic matter consists of two major components: bulge and disks. The bulge component is characterized by the Hernquist potential \citep{deSalas}:
\begin{equation}
\Phi=-\frac{GM_b}{r+r_b}.
\end{equation}
The values of the constant parameters $r_b$ and $M_b$ are shown in Table 1. The corresponding velocity contribution is given by
\begin{equation}
V_{\rm bulge}^2=r\frac{d\Phi}{dr}=\frac{GM_br}{(r+r_b)^2}.
\end{equation}
For the disks component, it consists of a stellar disk, two dust disks (cold and warm dust), and two gas disks (H$_2$ and HI gas). All of the surface mass density in the disk components can be represented by an exponential function \citep{Misiriotis}:
\begin{equation}
\Sigma=\Sigma_0 \exp \left(-\frac{R}{R_0} \right),
\end{equation}
where $\Sigma_0=M_d/(2\pi R_0^2)$ is the central surface density. The corresponding scale disk radius $R_0$ and disk mass $M_d$ for each disk component can be found in Table 1. The RC contribution for each disk is given by \citep{Freeman}
\begin{eqnarray}
V_{\rm disk}^2&=& \frac{GM_dR^2}{2R_0^3} \left[I_0 \left(\frac{R}{2R_0} \right)K_0 \left(\frac{R}{2R_0} \right) \right. \nonumber\\
&& \left. -I_1 \left(\frac{R}{2R_0} \right)K_1 \left(\frac{R}{2R_0} \right) \right],
\end{eqnarray}
where $I_n$ and $K_n$ are the modified Bessel functions of the $n^{\rm th}$ kind.

The baryonic RC contributed by the bulge and disks components is given by
\begin{equation}
V_{\rm bar}^2=V_{\rm bulge}^2+\sum V_{\rm disk}^2.
\end{equation}
Therefore, we can get the baryonic radial acceleration:
\begin{equation}
a_{\rm bar}=\frac{V_{\rm bar}^2}{R}.
\end{equation}

\section{Data analysis}
By using two samples of GRCs obtained in \citet{Ou,Jiao}, we plot the values of $a_{\rm dyn}$ against $a_{\rm bar}$ in Fig.~1, which represent the GRARs. We can see that both GRARs are consistent with each other. However, when comparing the GRARs with the RAR obtained by using SPARC data (see the brown error bars in Fig.~1), we can see some systematic deviation between them. The two regions of the GRARs lie near the edge of the $1\sigma$ margins of the SPARC RAR across $\log a_{\rm bar}=-10.75$ to $\log a_{\rm bar}=-9.75$. When $\log a_{\rm bar}<-10.75$, a significant drop in $a_{\rm dyn}$ can be seen when $a_{\rm bar}$ decreases. This behavior is mainly related to the GRC decline at $R>20$ kpc \citep{Ou,Jiao}. However, the SPARC RAR does not show any significant drop in $a_{\rm dyn}$ when $\log a_{\rm bar}\sim -10$ \citep{McGaugh}. If we look at the entire range of the latest study of the RAR, the drop in $a_{\rm dyn}$ occurs at $\log a_{\rm bar}<-14$, but not at $\log a_{\rm bar}<-10.75$ \citep{Mistele}. Nevertheless, since the uncertainty of the SPARC RAR is relatively large, such a deviation between the regions of the GRARs and SPARC RAR is still statistically allowed. 

Nevertheless, past studies claimed that there exists a universal analytic form of RAR like Eq.~(1) satisfying the SPARC data \citep{McGaugh,Mistele}. The best-fit acceleration scale is $a_0=(1.20 \pm 0.26) \times 10^{-10}$ m/s$^2$ \citep{McGaugh}. The blue dashed line in Fig.~1 indicates the best-fit analytic RAR with $a_0=1.2\times 10^{-10}$ m/s$^2$. If it represents the universal form of RAR, it significantly deviates from the GRAR. Following the same form of Eq.~(1), we can get the range of $a_0$ which can best fit with the GRAR. The best-fit acceleration scale for the RC samples in \citet{Jiao} and \citet{Ou} are $a_0=1.67\times 10^{-10}$ m/s$^2$ ($\chi^2=30.2$) and $a_0=1.88\times 10^{-10}$ m/s$^2$ ($\chi^2=403$) respectively. However, these best-fit values are ruled out by the corresponding GRC data at more than $2\sigma$ and $5\sigma$ respectively, which means that these so-called best-fit scenarios are indeed very poor fits. For the RC sample in \citet{Jiao}, the $5\sigma$ range is $a_0=1.67^{+0.25}_{-0.22} \times 10^{-10}$ m/s$^2$, which lies outside the best-fit range of $a_0$ in the SPARC RAR. Therefore, both GRCs show discrepancies with the universal form of RAR at more than $5\sigma$. 

As the analysis in \citet{Ou} has provided a more complete set of GRC sample with $R=6-27$ kpc, we further investigate the discrepancy statistically by the $\chi^2$ function using this sample. The corresponding statistical framework and the probability conversion can be found in \citet{Boudaud}. In Fig.~2, we plot the $\chi^2$ value against the acceleration scale $a_0$ by assuming that Eq.~(1) is the universal analytic form of RAR. We can see that the $\chi^2$ value depends sensitively on $a_0$. The best-fit $a_0=1.88\times 10^{-10}$ m/s$^2$ is ruled out far beyond $10\sigma$. For the range of $a_0=(1.20 \pm 0.26)\times 10^{-10}$ m/s$^2$ derived from the SPARC sample, the $\chi^2$ values are larger than 7000. Therefore, if Eq.~(1) is the best-fit analytic form satisfying the SPARC data, it would not satisfy the GRAR derived from \citet{Ou} for any $a_0$. In other words, it is quite likely that there is no universal RAR satisfying both GRC and SPARC data simultaneously.

\begin{figure}
\vskip 10mm
 \includegraphics[width=140mm]{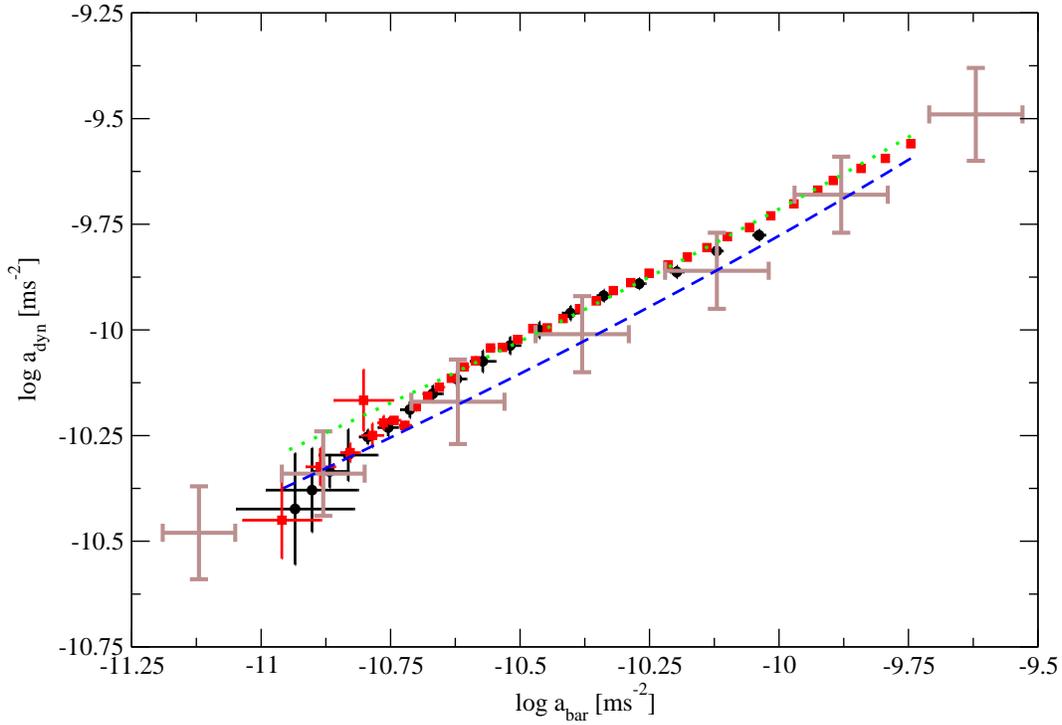}
 \caption{The black and red dots with error bars represent the GRAR derived from the GRC samples in \citet{Jiao} and \citet{Ou} respectively. The brown error bars indicate the uncertainties of the SPARC RAR obtained in \citet{McGaugh}. The blue dashed line represents the best-fit universal RAR using Eq.~(1) with $a_0=1.2\times 10^{-10}$ m/s$^2$. The green dotted line represents the best-fit GRAR using Eq.~(1) with $a_0=1.88 \times 10^{-10}$ m/s$^2$.}
\vskip 10mm
\end{figure}

\begin{figure}
\vskip 10mm
 \includegraphics[width=140mm]{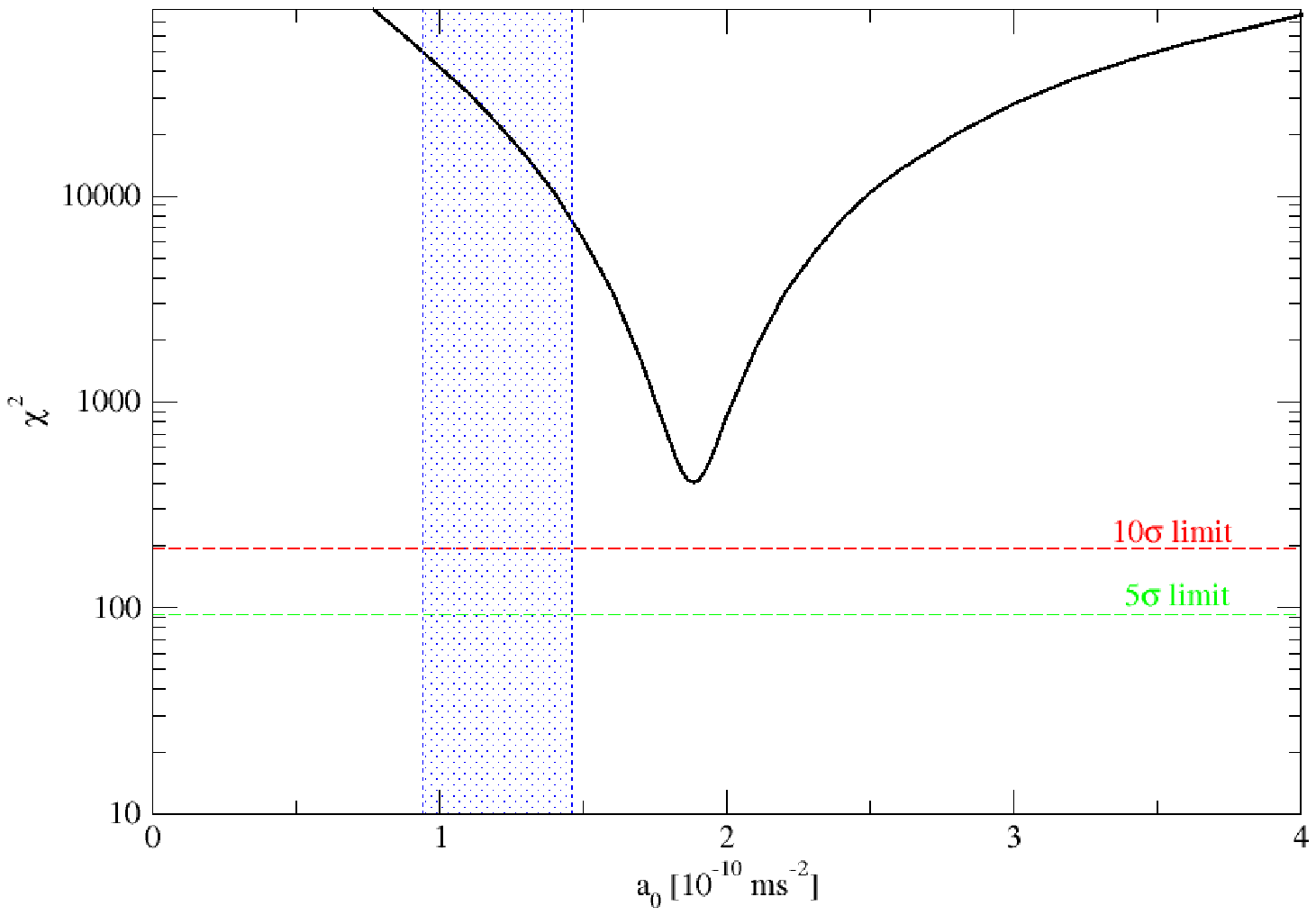}
 \caption{The black solid line represents the $\chi^2$ values against $a_0$. The green and red dashed lines indicate the upper limits of $\chi^2$ ruling out Eq.~(1) to account for the GRAR derived from the data in \citet{Ou} at $5\sigma$ and $10\sigma$ respectively. The region bounded by the blue dotted lines represents the best-fit range of $a_0$ derived from the SPARC sample \citep{McGaugh}.}
\vskip 10mm
\end{figure}

\begin{table}
\caption{The parameters of the baryonic model \citep{Misiriotis,deSalas,Jiao}.}

\begin{tabular}{ |l|l|l|}
 \hline\hline
  Component & $r_b$ or $R_0$ (kpc) & $M_b$ or $M_d$ ($M_{\odot}$) \\
  \hline
  Bulge & $0.700$ &  $1.550 \times 10^{10}$ \\
  Stellar disk & $2.35$ & $3.65 \times 10^{10}$ \\
  Cold dust disk & $5.0$ & $7.0 \times 10^7$ \\
  Warm dust disk & $3.3$ & $2.2\times 10^5$ \\
  H$_2$ gas disk & $2.57$ & $1.3\times 10^9$ \\ 
  HI gas disk & $18.24$ & $8.2\times 10^9$ \\
 \hline\hline
\end{tabular}
\end{table}

\section{Discussion}
Strictly speaking, this is not the first time to get the GRAR. A few recent studies have indeed considered or generated the GRAR by using the old data of the GRC \citep{Islam,Oman}. However, the major objective of using the GRAR in these studies is to differentiate different modified gravity theories or constrain the slope of the RAR, but not testing the existence of a universal RAR in galaxies. Moreover, some other studies claimed that the data of GRC is consistent with the RAR \citep{McGaugh2,McGaugh3}. However, the data of GRC used in these studies are not the updated one. In our study, we follow the data of the latest Gaia DR 3 which has given much better improved parallaxes and proper motions measurement than that in previous measurements \citep{Jiao}. Therefore, this is the first time to examine whether there exists a universal and consistent RAR simultaneously satisfying the latest high quality GRC data from Gaia DR 3 and the data in SPARC galaxies. We have shown that the latest GRC data from two different samples show a significant deviation from the claimed universal SPARC RAR. For a more complete sample of RC in \citet{Ou} (with a larger range of $R$), the best-fit $a_0$ is ruled out at more than $10\sigma$. Although the GRAR considered contains the data of one galaxy (i.e. our Galaxy) only, the GRC data used are robust and contain extremely small systematic and observational uncertainties \citep{Ou,Jiao}. Therefore, the resultant GRARs obtained are very reliable and we can arrive at a strong conclusion. If our Galaxy is not a special one, this implies that either there is no universal form of RAR or the so-called acceleration scale $a_0$ is not a universal constant. 

The existence of a universal RAR is a strong indicator of the correlation between dark matter and baryons. This provides a strong evidence to support modified gravity theories rather than the postulation of dark matter \citep{McGaugh,Li2}. If there is no universal RAR in galaxies, such a potential strong correlation between baryonic matter and dark matter may not exist. This also gives negative impact to any modified gravity theories which predict the association between dynamical mass and baryonic mass. In fact, some previous studies have already challenged that there is no universal form of RAR or there is no universal acceleration scale. For example, the RARs obtained by the data of galaxy clusters and elliptical galaxies are significantly different from the SPARC RAR \citep{Chan,Chan3}. Also, the best-fit values of $a_0$ are much larger than that obtained by SPARC data \citep{Chan}. Even using the SPARC sample, some studies argue that the universal acceleration scale does not exist \citep{Rodrigues}. Although these results are still controversial and inconclusive, our results add strong evidence to support falsifying the existence of a universal RAR, even in galaxies.

One may argue that the baryonic model used might not be accurate enough so that the discrepancy between the GRAR and the SPARC RAR is not justified. In fact, there are some other baryonic models proposed which might give different baryonic mass distributions \citep{deJong,Calchi}. Different baryonic models might follow similar functional forms of baryonic density distributions but with slightly different parameters. However, as shown in Fig.~1, the discrepancy originates from the fact that the baryonic mass in our Galaxy is smaller than that predicted by the SPARC RAR. In other words, any baryonic model which can give a larger baryonic mass might be able to alleviate the discrepancy. Based on the comprehensive analysis in \citet{Jiao} considering several most representative baryonic models, the baryonic model we used (called B2 model in \citet{Jiao}) has already given the largest amount of baryonic mass. The other baryonic models give 1\%-10\% smaller in baryonic mass (see Table 4 in \citet{Jiao}). Therefore, our baryonic model already provides the most optimal calculations for the GRAR to match the SPARC RAR. Roughly speaking, in order to match the best-fit SPARC RAR, the baryonic mass has to be larger by more than 30\%. In other words, appealing to the variation of baryonic models does not help in alleviating the discrepancy. 

As shown in Fig.~1 that the discrepancy between GRAR and SPARC RAR becomes much more significant in the small $a_{\rm bar}$ regime, we anticipate that more data at large $R$ (i.e. small $a_{\rm bar}$) can further verify our conclusion and test for dark matter and modified gravity theories \citep{Mercado}. In fact, there are some GRC data obtained for $R>30$ kpc using different methods \citep{Vasiliev,Wang}. However, the involved uncertainties are still quite large so that these data are not good enough for performing accurate analysis. To conclude, our analysis shows that either there is no universal form of RAR or the universal acceleration scale does not exist.

\section{Acknowledgements}
The work described in this paper was partially supported by a grant from the Research Grants Council of the Hong Kong Special Administrative Region, China (Project No. EdUHK 18300922).

\end{document}